# Relation between the Nusselt and Bejan numbers in natural convection

Takuya Masuda (益田 卓哉),[1,2,a] Toshio Tagawa (田川 俊夫)[2]

[1] Department of Integrated Engineering, National Institute of Technology, Yonago College,

4448 Hikona-cho, Yonago, Tottori 683-8502, Japan

[2] Department of Aeronautics and Astronautics, Tokyo Metropolitan University,

6-6 Asahigaoka, Hino, Tokyo 191-0065, Japan

[a] Author to whom correspondence should be addressed: masuda@yonago-k.ac.jp

ABSTRACT

This study derives a scaling law connecting the Nusselt ($Nu$) and Bejan ($Be$) numbers in natural convection. Combining entropy generation analysis with boundary-layer scaling, the relation $Be^{-1} - 1 = a\, Nu^b$ naturally emerges without explicit dependence on geometry or boundary conditions. This is achieved within the present scaling framework when transport is governed by a single control parameter. Numerical validation against several cases corroborates this scaling. This finding reveals a direct, quantitative link between heat transfer efficiency and thermodynamic irreversibility, suggesting a potentially universal constraint that governs convective transport.

Natural convection is considered as one of the fundamental transport mechanisms in fluid mechanics, driven predominantly by buoyancy forces. The associated heat transfer performance is conventionally characterized by the Nusselt number $Nu$, which exhibits power-law scaling with the Rayleigh number $Ra$: $Nu \sim Ra^n$.[1–3] This behavior has been confirmed across a diverse array of configurations, including Rayleigh–Bénard convection,[1,2] differentially heated cavities,[4,5] and horizontal convection.[6] Specifically, thermodynamic irreversibility in convective systems is extensively investigated through entropy generation analysis.[7–13] The Bejan number $Be$, defined as the ratio of heat-conduction-induced entropy generation to total entropy generation, serves as an indicator of the relative importance of thermal and viscous dissipation mechanisms.. Although $Nu$ and $Be$ are extensively used, traditionally, they have been treated as independent quantities. While numerous studies have examined their dependence on parameters such as $Ra$,[14–23] a direct functional relationship linking the two has not been formally identified.

A recent three-dimensional numerical investigation of thermomagnetic convection under a quadrupole magnetic field unveiled the following relation:[24]

$$Be^{-1} - 1 = a\,Nu^b, \qquad (1)$$

where $a$ and $b$ are constants. It should be noted that this relation was derived without explicitly involving geometric parameters or boundary conditions; instead, it emerges from the inherent scaling structure of entropy generation. This finding raises the possibility that Eq. (1) potentially applies to any natural convection system in which $Nu$ exhibits power-law dependence on a single control parameter such as $Ra$.

The purpose of this Letter is to formalize this concept and show that the relation given by Eq. (1) may constitute a general feature of natural convection that does not explicitly exhibit dependence on geometry or boundary conditions within the present scaling framework. Furthermore, this claim has been demonstrated through several cases, including a canonical benchmark problem: natural convection in a side-heated square cavity.[4,5]

The nondimensional governing equations take the following form:

$$\nabla \cdot \mathbf{U} = 0, \qquad (2)$$

$$\frac{D\mathbf{U}}{DT} = -\nabla P + Pr\nabla^2 \mathbf{U} - PrRa\Theta\mathbf{e}_y, \tag{3}$$

$$\frac{D\Theta}{DT} = \nabla^2\Theta. \tag{4}$$

The total entropy generation within the domain comprises thermal and viscous contributions:

$$s = s_\theta + s_\psi. \tag{5}$$

The dimensionless local entropy generation rates arising from heat transfer and fluid friction, $S_{\Theta,loc}$ and $S_{\Psi,loc}$, respectively, are expressed as:

$$S_{\Theta,loc} = |\nabla\Theta|^2, \tag{6}$$

$$S_{\Psi,loc} = 2\varphi \mathbf{D} : \mathbf{D}, \tag{7}$$

where

$$\mathbf{D} = \frac{1}{2}(\nabla\mathbf{U} + (\nabla\mathbf{U})^T). \tag{8}$$

Following previous studies,[9,11] the irreversibility distribution ratio is assigned $\varphi = 10^{-4}$. The domain-integrated entropy generation rates are obtained by integrating the local contributions over the entire computational domain. Accordingly, the Bejan number is defined as:

$$Be = \frac{S_\Theta}{S_\Theta + S_\Psi}. \tag{9}$$

Thus,

$$\frac{1}{Be} - 1 = \frac{S_\Psi}{S_\Theta}. \tag{10}$$

This expression indicates that $Be^{-1} - 1$ directly represents the ratio of viscous to thermal irreversibility.

The scaling behavior of the Nusselt number and entropy generation can be rationalized through boundary-layer arguments. In natural convection, the thermal boundary-layer thickness $\delta_\theta$ follows a power-law dependence on the governing parameter, reflecting the balance between buoyancy-driven flow and diffusive transport. Since the Prandtl number is fixed in the present study, the ratio between the momentum boundary-layer thickness $\delta_u$ and $\delta_\theta$ remains approximately constant. The Nusselt number scales as $Nu \sim l_c/\delta_\theta = \delta_\Theta^{-1}$. It follows that

$$Nu \sim Ra^n. \tag{11}$$

Given that the temperature gradient in the boundary layer scales as $|\nabla\Theta| \sim \delta_\Theta^{-1}$, thermal entropy generation, dominated by the boundary-layer contribution, obeys

$$S_\Theta \sim Ra^p. \tag{12}$$

Similarly, viscous entropy generation is governed by near-wall shear layers, where the velocity gradient scales as $|\nabla \mathbf{U}| \sim U_c/\delta_U$. Consequently,

$$S_\Psi \sim Ra^q. \tag{13}$$

Therefore, $Be^{-1} - 1 = a\, Nu^b$ holds up to multiplicative constants. Since this derivation does not contain any geometric or boundary-condition-specific parameters, it suggests a degree of generality.

The proposed relation is illustrated by considering natural convection in a square cavity with differentially heated vertical walls, a standard benchmark problem widely employed for computational fluid dynamics validation. This configuration has been extensively studied since the pioneering work of De Vahl Davis[4] with well-established flow and heat transfer characteristics. The numerical setup consists of a square cavity filled with air ($Pr = 0.71$). The left and right vertical walls are maintained at constant high and low temperatures, respectively, with the horizontal walls assumed to be adiabatic. The governing equations are solved using a second-order accurate central difference scheme with finite-volume method, and steady-state solutions are obtained for Rayleigh numbers in the range $10^3 \leq Ra \leq 10^7$.

The results are presented in Fig. 1. Consistent with classical studies, the Nusselt number follows $Nu \sim Ra^n$. Entropy generation is evaluated by integrating thermal and viscous contributions over the domain. The resulting data confirm that $Be^{-1} - 1 = a\,Nu^b$ holds with high accuracy across the entire $Ra$ range examined in the present configuration.

Furthermore, the applicability of the proposed correlation to different systems was examined. First, as shown in Fig. 2(a), the Prandtl number in the benchmark case was varied to $Pr = 7.1$ and $0.1$. For the unsteady regime ($Pr = 0.1$ and $Ra = 10^7$), time-averaged values were employed. Next, as shown in Fig. 2(b), Rayleigh–Bénard convection in a square enclosure was computed. As can be seen from Fig. 2(d), the behavior near the critical point was also properly reproduced. In addition, as shown in Fig. 2(c), a concentric double-cylinder configuration was considered, where the inner cylinder was heated and the outer cylinder was cooled, with a radius ratio of 2. In all cases, the relation given by Eq. (1) was found to hold well.

Despite the pronounced dependence of flow structure on $Ra$, spanning the transition from conduction-dominated to convection-dominated regimes, the $Nu - Be$ relation remains unchanged. Consequently, this confirms that the relation reflects a fundamental balance between heat transport and irreversibility, independent of the details of flow structure.

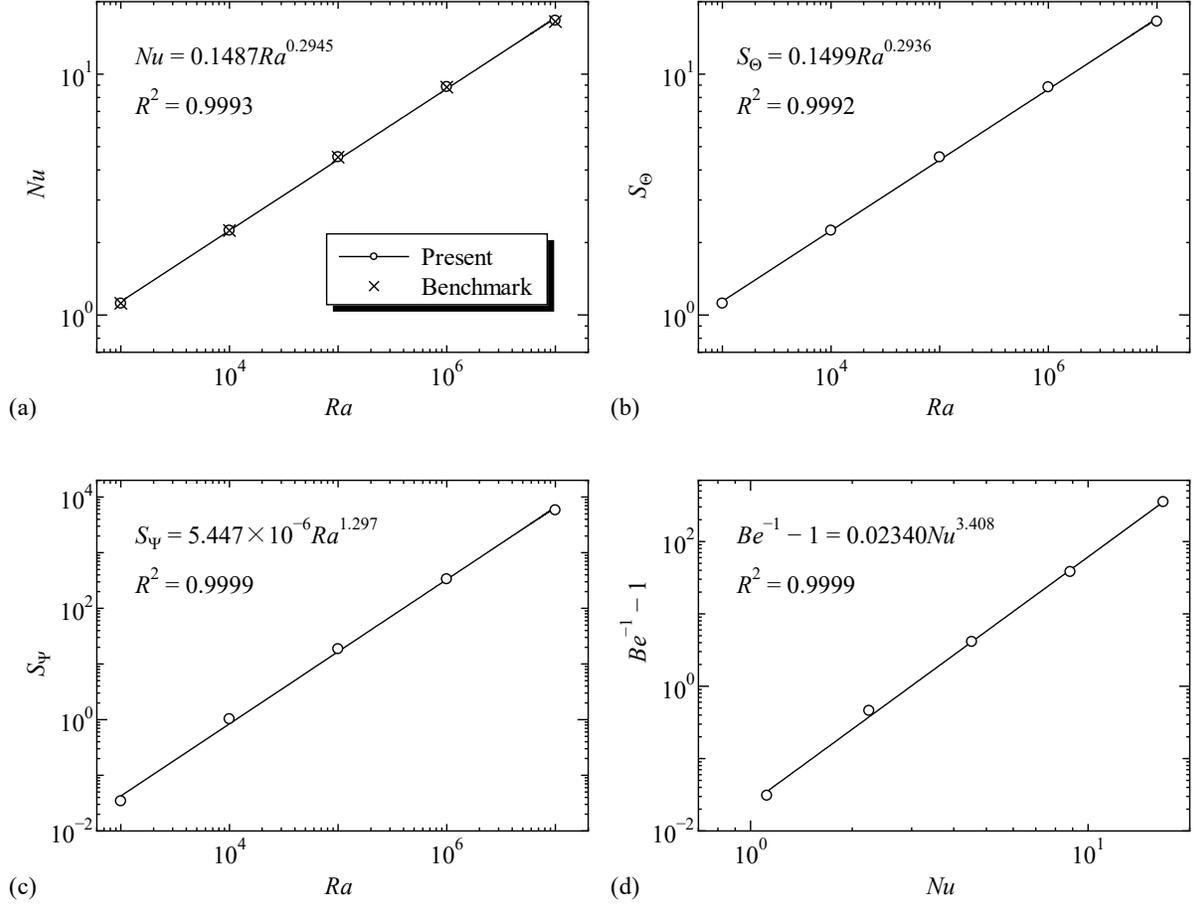

FIG. 1. (a) Average Nusselt number $Nu$, (b) dimensionless total entropy generation due to heat conduction $S_\Theta$, and (c) dimensionless total entropy generation due to viscous dissipation $S_\Psi$ as functions of the Rayleigh number $Ra$, together with (d) $Be^{-1} - 1$ plotted as a function of $Nu$. Here, $R^2$ denotes the coefficient of determination. Reference values of $Nu$ are taken from de Vahl Davis[4] for $Ra = 10^3$–$10^5$ and from Le Quéré[5] for $Ra = 10^6$–$10^7$.

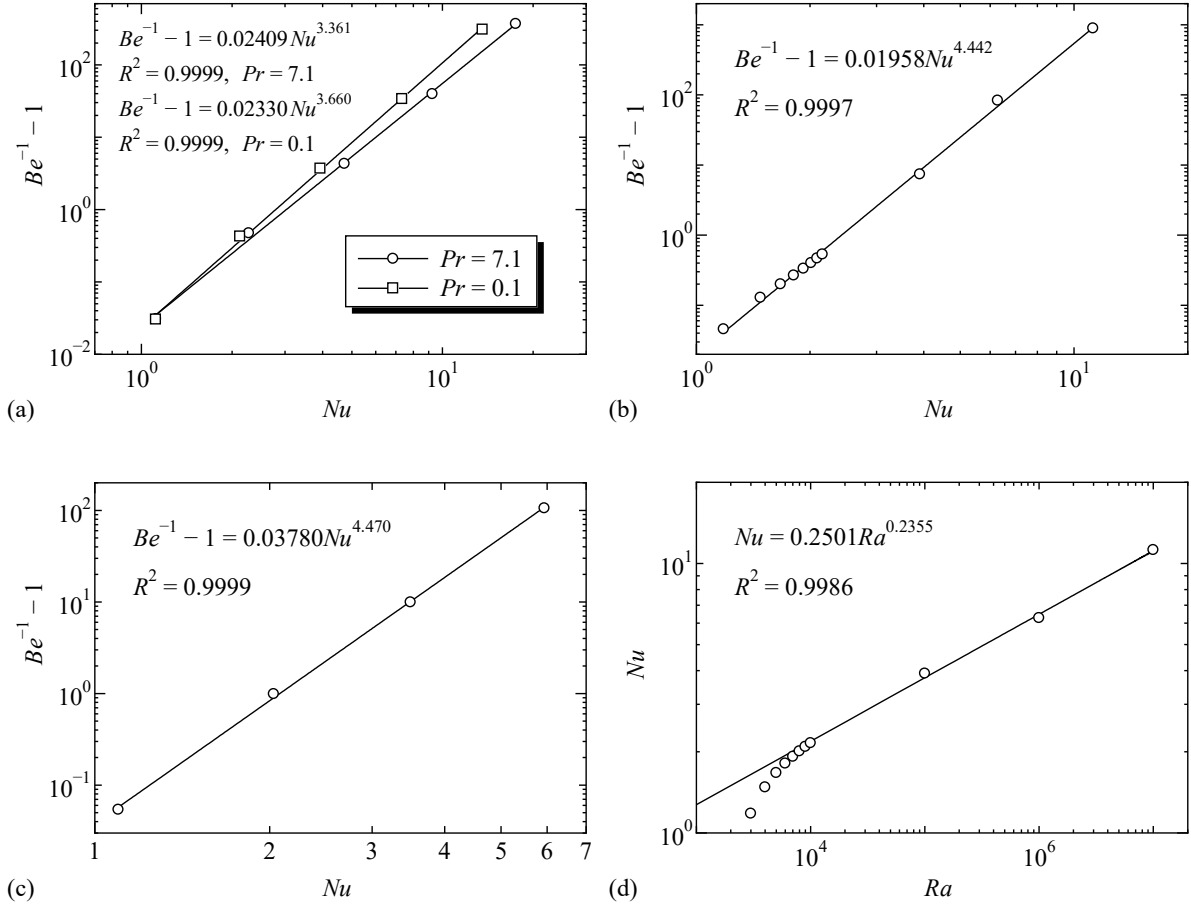

FIG. 2. Distribution of $Be^{-1} - 1$ as a function of $Nu$: (a) variation with $Pr$ for the benchmark case; (b) Rayleigh–Bénard convection in a square cavity ($3000 \leq Ra \leq 10^7$); and (c) concentric double-cylinder geometry ($10^3 \leq Ra \leq 10^6$). In addition, (d) shows the variation of $Nu$ with $Ra$ for Rayleigh–Bénard convection. The fitted functions in the figure are obtained over the range $10^4 \leq Ra \leq 10^7$.

ACKNOWLEDGMENTS

This work was supported by JSPS KAKENHI Grant Number JP25K17546.

REFERENCES

1. W. V. Malkus, "The heat transport and spectrum of thermal turbulence," Proc. R. Soc. Lond., Ser. A **225**(1161), 196–212 (1954).
https://doi.org/10.1098/rspa.1954.0197

2. L. N. Howard, "Heat transport by turbulent convection," J. Fluid Mech **17**(3), 405–432 (1963).
https://doi.org/10.1017/S0022112063001427

3. S. Grossmann and D. Lohse, "Scaling in thermal convection: a unifying theory," J. Fluid Mech **407**, 27–56 (2000).
https://doi.org/10.1017/S0022112099007545

4. G. de Vahl Davis, "Natural convection of air in a square cavity: a bench mark numerical solution," Int. J. Numer. Methods Fluids **3**(3), 249–264 (1983).
https://doi.org/10.1002/fld.1650030305

5. P. Le Quéré, "Accurate solutions to the square thermally driven cavity at high Rayleigh number," Comput. Fluids **20**(1), 29–41 (1991).
https://doi.org/10.1016/0045-7930(91)90025-D

6. H. T. Rossby, "On thermal convection driven by non-uniform heating from below: an experimental study," Deep-Sea Res. Oceanogr. Abstr. **12**(1), 9–16 (1965).
https://doi.org/10.1016/0011-7471(65)91336-7

7. A. Bejan, "A study of entropy generation in fundamental convective heat transfer," J. Heat Transfer. **101**(4), 718–725 (1979).
https://doi.org/10.1115/1.3451063

8. B. S. Yilbas, S. Z. Shuja, S. A. Gbadebo, H. I. Abu Al-Hamayel, and K. Boran, "Natural convection and entropy generation in a square cavity," Int. J. Energy Res. **22**(14), 1275–1290 (1998).
https://doi.org/10.1002/(SICI)1099-114X(199811)22:14%3C1275::AID-ER453%3E3.0.CO;2-B

9. M. Magherbi, H. Abbassi, and A. Ben Brahim, "Entropy generation at the onset of natural convection," Int. J. Heat Mass Transf. **46**(18), 3441–3450 (2003).
https://doi.org/10.1016/S0017-9310(03)00133-9

10. S. Mahmud and R. A. Fraser, "Magnetohydrodynamic free convection and entropy generation in a square porous cavity," Int. J. Heat Mass Transf. **47**(14–16), 3245–3256 (2004).
https://doi.org/10.1016/j.ijheatmasstransfer.2004.02.005

11  G. G. Ilis, M. Mobedi, and B. Sunden, "Effect of aspect ratio on entropy generation in a rectangular cavity," Int. Commun. Heat Mass Transf. **35**(6), 696–703 (2008).
https://doi.org/10.1016/j.icheatmasstransfer.2008.02.002

12  R. D. C. Oliveski, M. Macagnan, and J. B. Copetti, "Entropy generation and natural convection in rectangular cavities," Appl. Therm. Eng. **29**(8–9), 1417–1425 (2009).
https://doi.org/10.1016/j.applthermaleng.2008.07.012

13  H. F. Oztop and K. Al-Salem, "A review on entropy generation in natural and mixed convection heat transfer for energy systems," Renew. Sustain. Energy Rev. **16**(1), 911–920 (2012).
https://doi.org/10.1016/j.rser.2011.09.012

14  P. Mayeli and G. J. Sheard, "An entropy generation analysis of horizontal convection under the centrifugal buoyancy approximation," Int. Commun. Heat Mass Transfer **133**, 105923 (2022).
https://doi.org/10.1016/j.icheatmasstransfer.2022.105923

15  B. Iftikhar, T. Javed, and M. A. Siddiqu, "Entropy generation analysis during MHD mixed convection flow of non-Newtonian fluid saturated inside the square cavity," J. Comput. Sci. **66**, 101907 (2023).
https://doi.org/10.1016/j.jocs.2022.101907

16  S. Alqaed, F. A. Almehmadi, M. Sharifpur, and J. Mustafa, "Study of Bejan number and entropy generation for mixed convection of nanofluid flow inside a chamber under an inclined magnetic field," J. Magn. Magn. Mater. **594**, 171849 (2024).
https://doi.org/10.1016/j.jmmm.2024.171849

17  A. Punia and R. K. Ray, "New higher-order super-compact finite difference scheme to study three-dimensional natural convection and entropy generation in power-law fluids," Phys. Fluids **37**(1), 013324 (2025).
https://doi.org/10.1063/5.0246131

18  B. Souayeh, "Influence of boundary condition variations on magnetohydrodynamics natural convection and entropy generation in a ternary nanofluid filled-square cavity with elliptic cylinder," Phys. Fluids **37**(4), 042009 (2025).
https://doi.org/10.1063/5.0269514

19  S. Hansda and S. Soren, "Entropy generation and thermosolutal convection in a radiative porous chamber filled with a Casson-based ternary hybrid nanofluid and a cold square obstacle," Phys. Fluids **37**(8), 082058 (2025).
https://doi.org/10.1063/5.0281788

20  R. Ali, R. Imtiaz, Z. Badshah, A. Ali, and A. S. Hendy, "Finite element analysis of natural convection and entropy generation under the influence of magnetic field in a wavy enclosure,"

Phys. Fluids **37**(8), 083620 (2025).
https://doi.org/10.1063/5.0284120

21  R. Aich, D. Bhargavi, and K. Vajravelu, "Development of magnetothermal convection through an anisotropic porous channel with entropy generation," Phys. Fluids **37**(8), 083626 (2025).
https://doi.org/10.1063/5.0287837

22  Y. U. B. Turabi, S. Ahmad, and S. Ahmad, "Finite element and neural network modeling of thermal energy storage and entropy behavior in a wavy porous triangular enclosure with nano-encapsulated phase change materials," Phys. Fluids **37**(10), 102025 (2025).
https://doi.org/10.1063/5.0296914

23  G. Ali, Z. Ali, and H. A. Ghazwani, "Entropy generation and Bejan number optimization in fractional dusty nanofluid of free convectional flow," J. Korean Phys. Soc. **88**(5), 593–605 (2026).
https://doi.org/10.1007/s40042-025-01524-1

24  T. Masuda and T. Tagawa, "Three-dimensional numerical simulation of natural convection and entropy generation in rectangular enclosure under magnetic quadrupole field," submitted.